# Social distancing and epidemic resurgence in agent-based Susceptible-Infectious-Recovered models

Ruslan I. Mukhamadiarov[1], Shengfeng Deng[1,2], Shannon R. Serrao[1], Priyanka[1], Riya Nandi[1], Louie Hong Yao[1], and Uwe C. Täuber*[1,3]

[1] Department of Physics and Center for Soft Matter and Biological Physics, Virginia Tech, Blacksburg, VA 24061, USA.

[2] Key Laboratory of Quark and Lepton Physics (MOE) and Institute of Particle Physics, Central China Normal University, Wuhan 430079, China.

[3] Faculty of Health Sciences, Virginia Tech, Blacksburg, VA 24061, USA.

* Uwe C. Täuber
**Email:** tauber@vt.edu

All authors contributed equally to this work

## ABSTRACT

Once an epidemic outbreak has been effectively contained through non-pharmaceutical interventions, a safe protocol is required for the subsequent release of social distancing restrictions to prevent a disastrous resurgence of the infection. We report individual-based numerical simulations of stochastic susceptible-infectious-recovered model variants on four distinct spatially organized lattice and network architectures wherein contact and mobility constraints are implemented. We robustly find that the intensity and spatial spread of the epidemic recurrence wave can be limited to a manageable extent provided release of these restrictions is delayed sufficiently (for a duration of at least thrice the time until the peak of the unmitigated outbreak) and long-distance connections are maintained on a low level (limited to less than five percent of the overall connectivity).

## Introduction

The COVID-19 pandemic constitutes a severe global health crisis. Many countries have implemented stringent non-pharmaceutical control measures that involve behavioral change of the public such as social distancing, using face-coverings and mobility reduction enforced by lockdowns in their populations. This has led to remarkably successful deceleration and significant `flattening of the curve' of the infection outbreaks, albeit at tremendous economic and financial costs (1, 2). At this point, societies are in dire need of designing a secure (partial) exit strategy wherein the inevitable recurrence of the infection among the significant non-immune fraction of the population can be thoroughly monitored with sufficient spatial resolution and reliable statistics, provided that dependable, frequent, and widespread virus testing capabilities are accessible and implemented. Until an effective and safe vaccine is widely available, this would ideally allow the localized implementation of rigorous targeted disease control mechanisms that demonstrably protect people's health while the paralyzed branches of the economy are slowly rebooted.

Mathematical analysis and numerical simulations of infection spreading in generic epidemic models are crucial for testing the efficacy of proposed mitigation measures, and the timing and pace of their gradual secure removal. Specifically, the employed mathematical models need to be (i) stochastic in nature in order to adequately account for randomly occurring or enforced disease extinction in small isolated communities, as well as for rare catastrophic infection boosts and (ii) spatially resolved such that they properly capture the significant emerging correlations among the susceptible and immune subpopulations. These distinguishing features are notably complementary to the more detailed and comprehensive computer models utilized by researchers at the University of Washington, Imperial College London, the Virginia Bioinformatics Institute, and others: see, e.g., (3-8).

We report a series of detailed individual-based Monte Carlo computer simulation studies for stochastic variants (9,10) of the paradigmatic Susceptible-Infectious-Recovered (*SIR*) model (11,12) for a community of about 100,000 individuals. To determine the robustness of our results and compare the influence of different contact characteristics, we ran our stochastic model on four distinct spatially structured architectures, namely i) regular two-dimensional square lattices, wherein individuals move slowly and with limited range, i.e., spread diffusively; ii) two-dimensional small-world networks that in addition incorporate substantial long-distance interactions and contaminations; and finally on iii) random as well as iv) scale-free social contact networks. For each setup, we investigated epidemic outbreaks with model parameters informed by the known COVID-19 data (6,13). To allow for a direct comparison, we extracted the corresponding effective infection and recovery rates by fitting the peak height and the half-peak width of the infection growth curves with the associated classical deterministic SIR rate equations that pertain to a well-mixed setting. We designed appropriate implementations of social distancing and contact reduction measures on each architecture by limiting or removing connections between individuals. This approach allowed us to generically assess the efficacy of non-pharmaceutical control measures.

Although each architecture entails varied implementations of social distancing measures, we find that they all robustly reproduce both the resulting reduced outbreak intensity and growth speed. As anticipated, a dramatic resurgence of the epidemic occurs when mobility and contact restrictions are released too early. Yet if stringent and sufficiently long-lasting social distancing measures are imposed, the disease may go extinct in the majority of isolated small population groups. In our spatially extended lattice systems, disease spreading then becomes confined to the perimeters of a few larger outbreak regions, where it can be effectively localized and specifically targeted. For the small-network architecture, it is however imperative that all long-range connections remain curtailed to a very low percentage for the control measures to remain effective. Intriguingly, we observe that an infection outbreak spreading through a static scale-free network effectively randomizes its connectivity for the remaining susceptible nodes, whence the second wave encounters a very different structure.

In the following sections, we briefly describe the methodology and algorithmic implementations as well as pertinent simulation results for each spatial or network structure; additional details are provided in the Supplementary Materials. We conclude with a comparison of our findings and a summary of their implications.

# Results

**Square lattices with diffusive spreading**

Our first architecture is a regular two-dimensional square lattice with linear extension $L = 448$ subject to periodic boundary conditions (i.e., on a torus). Initially, $N = S(0) + I(0) + R(0) = 100,000$ individuals with fixed density $\rho = N/L^2 \approx 0.5$ are randomly placed on the lattice, with at most one individual allowed on each site. Almost the entire population begins in the susceptible state $S(0)$; we start with only 0.1 % infected individuals, $I(0) = 100$, and no recovered (immune) ones, $R(0) = 0$. We note that in stochastic simulations, random fluctuations often lead to the initial infectious population recovering so fast that the epidemic dies out before it can cause an outbreak; therefore we chose to seed the system with 100 randomly placed infected individuals. This initial configuration is moreover motivated by enforced lockdowns and travel restrictions which essentially stops the external influx of new infections. Subsequently, all individuals may move to neighboring empty lattice sites with diffusion rate $d$ (here we set this hopping probability to 1). Upon their encounter, infectious individuals irreversibly change the state of neighboring susceptible ones with set rate $r: S + I \rightarrow I + I$. Any infected individual spontaneously recovers to an immune state with fixed rate $a: I \rightarrow R$. (Details of the simulation algorithm are presented in the Supplementary Materials.) For the recovery period, we choose $1/a \cong 6.667$ days (1 day is equivalent to one Monte Carlo step, *MCS*) informed by known COVID-19 characteristics (13). To determine the infection rate $r$, we run simulations for various values, fit the peak height and width of the ensuing epidemic curves with the corresponding SIR rate equations to extract the associated basic reproduction ratio $R_0$ (as explained in the Supplementary Materials, see Figure S1), and finally select that value for $r$ for our individual-based Monte Carlo simulations that reproduces the $R_0 \approx 2.4$ for COVID-19 (6). We perform 100 independent simulation runs with these reaction rates, from which we obtain the averaged time tracks for $I(t)$ and $R(t)$, while of course $S(t) = N - I(t) - R(t)$ and $R(t) = a \int_0^t I(t') \, dt'$.

The standard classical SIR deterministic rate equations assume a well-mixed population and constitute a mean-field type of approximation wherein stochastic fluctuations and spatial as well as temporal correlations are neglected; see, e.g., (14,15). Near the peak of the epidemic outbreak, when many individuals are infected, this description is usually adequate, albeit with coarse-grained `renormalized' rate parameters that effectively incorporate fluctuation effects at short time and small length scales. However, the mean-field rate equations are qualitatively insufficient when the infectious fraction $I(t)/N$ is small, whence both random number fluctuations and the underlying discreteness and associated internal demographic noise become crucial (14-16). Already near the epidemic threshold, which constitutes a continuous dynamical phase transition far from thermal equilibrium, c.f. Figure S2 in the Supplementary Materials, the dynamics is dominated by strong critical point fluctuations. These are reflected in characteristic initial power laws rather than simple exponential growth of the $I(t)$ and $R(t)$ curves (17), as demonstrated in Figure S1 (Supplemental Information).

Nor can the deterministic rate equations capture stochastic disease extinction events that may occur at random in regions where the infectious concentration has reached small values locally. The rate equations may be understood to pertain to a static and fully connected network; in contrast, the spreading dynamics on a spatial setting continually rewires any infectious links keeping the epidemic active (8,18). Consequently, once the epidemic outbreak threshold is exceeded, the *SIR* rate equations markedly underestimate the time-integrated outbreak extent reflected in the ultimate saturation level $R_\infty = R(t \to \infty)$, as is apparent in the comparison Figure S1 (Supplemental Information).

Once the instantaneous infectious fraction of the population has reached the threshold 10 %, $I(t) = 0.1 \, N$, we initiate stringent social distancing that we implement through a strong repulsive interaction between any occupied lattice sites (with $n_i = 1$), irrespective of their states $S$, $I$, or $R$; and correspondingly an attractive force between filled and empty ($n_i = 0$) sites, namely the repulsive interaction energy $V(\{n_i\}) = K \sum_{<i,j>} (2 \, n_i - 1) \, (2 \, n_j - 1)$ with dimensionless strength $K = 1$, where the sum extends only over nearest-neighbor pairs on the square lattice. The transfer of any individual from an occupied to an adjacent empty site is subsequently determined through the ensuing energy change $\Delta V$ by the Metropolis transition probability $w = min\{1, \, exp \, (-\Delta V)\}$ (19,20), which replaces the unmitigated hopping rate $d$. As a result, both the mobility as well as any direct contact between individuals on the lattice are quickly and drastically reduced. With this social distancing mechanism, our system effectively operates like an adaptive network (21), where all types of links, rather than only the *S-I* links (22), tend to be dynamically suppressed during the lockdown period. For sufficiently small total density $\rho = N/L^2$, most of the individuals eventually become completely isolated from each other. For our $\rho = 0.5$, the disease will continue to spread for a short period, until the repulsive potential has induced sufficient spatial anti-correlations between the susceptible individuals. The social-distancing interaction is sustained for a time duration $T$, and then switched off again.

*Figure 1 about here.*

Figure 1 depicts two sets of Monte Carlo simulation snapshots, each beginning at the moment when social distancing is switched on. The second column shows the configurations when the repulsive interaction $V$ is turned off again after respectively $T = 2/a$ (top), and $T = 10/a$ (bottom), while the last two sets of snapshots illustrate the subsequent resurgence of the outbreak *(\*)*. With increasing mitigation duration $T$, the likelihood for the disease to locally go extinct in isolated population clusters grows markedly. As seen in the bottom row, the prevalence and spreading of the infection thus becomes confined to the perimeters of a mere few remaining centers. Hence we observe drastically improved mitigation effects for extended $T$: As shown in Figure 2, the resurgence peak in the $I(t)$ curve assumes markedly lower values and is reached after much longer times. In fact, the time $\tau(T)$ for the infection outbreak to reach its second maximum increases exponentially with the social-distancing duration, as evidenced in the inset of Figure 2 (see also Figure 6 below). We emphasize that localized disease extinction and spatial confinement of the prevailing disease clusters represent correlation effects that cannot be captured in the *SIR* mean-field rate equation description.

*Figure 2 about here.*

**Two-dimensional small-world networks**

In modern human societies, individuals as well as communities feature long-distance connections that represent `express' routes for infectious disease spreading in addition to short-range links with their immediate neighbors. To represent this situation, we extend our regular lattice with diffusive propagation to a two-dimensional Newman-Watts small-world network (23), which was previously applied to the study of plant disease proliferation (24). Diffusive propagation is the manifestation of the natural movement of the individuals over the spatial extent of the lattice. In contrast to the Watts-Strogatz model (25), in which the small-world property is generated through rewiring bonds of a one-dimensional chain of sites, a Newman-Watts small-world network may be constructed as follows: For each nearest-neighbor bond, a long-distance link (or `short-cut') is added with probability $\varphi$ between randomly chosen pairs of vertices. As illustrated in Figure S3 (Supplemental Information), the resulting network features $2\varphi L^2$ long-distance links, with mean coordination number $<k> = 4(1+\varphi)$.

Again, each vertex may be in either of the states $S$, $I$, $R$, or empty, and each individual can hop to another site along any (nearest-neighbor or long-distance) link with a total diffusion rate $d$. While network graphs have been widely employed before to represent human social interactions, we emphasize that our approach substantially differs in that we simulate a fully stochastic set of SIR reactions on dynamically changing networks that have an underlying static small-network structure. A typical snapshot of the SIR model on this small-world architecture is shown in Figure S3 (Supplemental Information). The unmitigated simulation parameters are: $L = 1,000$, $N = 100,000$, $I(0) = 100$, $d = 1$, and $\varphi = 0.6$. The presence of long-range links increases the mean connectivity, rendering the population more mixed, which in turn significantly facilitates epidemic outbreaks (see Figure S4 in the Supplemental Information). We remark that for the *SIR* dynamics, the Newman-Watts small-world network effectively interpolates between a regular two-dimensional lattice and a scale-free network dominated by massively connected hubs; moreover, as the hopping probability $d \to 0$, the small-world network is effectively rendered static.

*Figure 3 about here.*

In the two-dimensional small-world network, we may introduce social-distancing measures through two distinct means: i) We can globally diminish mobility by adopting a reduced overall diffusion rate $d' < 1$; and/or ii) we can drastically reduce the probability of utilizing a long-distance connection to $d_\varphi \ll 1$. We have found that the latter mitigation strategy of curtailing the infection short-cuts into distant regions has a far superior effect. Therefore, in Figure 3 we display the resulting data for such a scenario where we set $d_\varphi = 0.05$, yet kept the diffusion rate unaltered at $d = 1$; as before, this control was triggered once $I(t) = 0.1 N$ had been reached in the course of the epidemic. The resurgence peak height and growth rate become even more stringently reduced with extended mitigation duration than for (distinct) social distancing measures implemented on the regular lattice.

**Random and scale-free contact networks**

Finally, we run the stochastic *SIR* dynamics on two different static structures, namely i) randomly connected and ii) scale-free contact networks. Each network link may be in either the $S$, $I$, or $R$ configurations, which are subject to the *SIR* reaction rules, but we do not allow movement among the network vertices. For the random network, we uniformly distribute 1,000,000 edges among $N = 100,000$ nodes; this yields a Poisson distribution for the connectivity with preset mean (equal to the variance) $<k> = (\Delta k)^2 = 20$. For the scale-free network, we employ the Barabasi-Albert graph construction (26), where each new node is added successively with $k = 4$ edges, to yield a total of 799,980 edges. The connectivity properties in these quite distinct architectures are vastly different, since the scale-free networks feature prominent `hubs' through which many other nodes are linked. In the epidemic context, these hubs represent super-spreader centers through which a large fraction of the population may become infected (10,27).

To implement the stochastic *SIR* dynamics on either contact network, we employ the efficient rejection-free Gillespie dynamical Monte Carlo algorithm: Each reaction occurs successively, but the corresponding time duration between subsequent events is computed from the associated probability function (28) (for details, see Supplemental Information). The random social network may be considered an emulation of the well-connected mean-field model. Indeed, we obtain excellent agreement for the temporal evolution of the *SIR* dynamics in these two systems with $a = 0.15$ *MCS* (for the scale-free network, a small adjustment to an effective mean-field recovery rate $a \approx 0.18$ *MCS* is required). A variety of measures can be taken to effectively control the epidemic spread on a network (21,22). We implement a `complete lockdown' mitigation strategy: Once the threshold $I(t) = 0.1\,N$ has been reached, we immediately cut all links for a subsequent duration $T$; during that time interval, only spontaneous recovery $I \to R$ can occur.

*Figure 4 about here.*

In Figure 4, we discern a markedly stronger impact of this lockdown on the intensity of the epidemic resurgence in both these static contact network architectures, see also Figure 6A below. On the other hand, the mitigation duration influences the second infection wave less strongly, with the time until its peak has been reached growing only linearly with $T$: $\tau(T) \sim T$, as is visible in Figure 6B. There is however a sharp descent in resurgent peak height beyond an apparent threshold $T > 7/a$ for the random network, and $T > 8/a$ for the scale-free network. For both the two-dimensional regular lattice and small-world structure, a similar sudden drop in the total number of infected individuals (Figure 6B) requires a considerably longer mitigation duration: In these dynamical networks, the repopulation of nodes with infective individuals facilitates disease spreading, thereby diminishing control efficacy.

We remark that if a drastically reduced diffusivity $d' \ll 1$ is implemented, the small-world results closely resemble those for a randomly connected contact network (Figure 6A).

*Figure 5 about here.*

Moreover, we have observed an unexpected and drastic effective structural change in the scale-free network topology as a consequence of the epidemic outbreak infecting its susceptible nodes. Naturally, the highly connected hubs are quickly affected, and through transitioning to the recovered state, become neutralized in further spreading the disease. As shown in Figure 5, as the infection sweeps through the network (in the absence of any lockdown mitigation), the distribution of the remaining active susceptible-infectious (*SI*) links remarkably changes from the initial scale-free power law with exponent $-1/2$ to a more uniform, almost randomized network structure. The disease resurgence wave thus encounters a very different network topology than the original outbreak.

# Discussion

In this study, we implemented social distancing control measures for simple stochastic *SIR* epidemic models on regular square lattices with diffusive spreading, two-dimensional Newman-Watts small-world networks that include highly infective long-distance connections, and static contact networks, either with random connectivity or scale-free topology. In these distinct architectures, all disease spreading mitigation measures, be that through reduced mobility and/or curtailed connectivity, must of course be implemented at an early outbreak stage, but also maintained for a sufficient duration to be effective. In Figure 6, we compare salient features of the inevitable epidemic resurgence subsequent to the elimination of social distancing restrictions, namely the asymptotic fraction $R_\infty/N$ of recovered individuals, i.e., the integrated number of infected individuals; and the time $\tau(T)$ that elapses between the release and the peak of the second infection wave, both as function of the mitigation duration $T$. We find that the latter grows exponentially with $T$ on both dynamical lattice architectures, but only linearly on the static networks (Figure 6B). Furthermore, as one would expect, the mean-field rate equations pertaining to a fully connected system describe the randomly connected network very well.

*Figure 6 about here.*

In stark contrast to the mean-field results (indicated by the purple lines in Figure 6), the data for the lattice and network architectures reveal marked correlation effects that emerge at sufficiently long mitigation durations $T$. For $T > 8/a$ in the static networks, and $T > 12/a$ in the lattice structures, the count of remaining infectious individuals

$I$ becomes quite low; importantly, these are also concentrated in the vicinity of a few persisting infection centers. This leads to a steep drop in $R_\infty/N$, the total fraction of ever infected individuals, by a factor of about 4 in the static network, and 3 in the dynamic lattice architectures. Thus, in these instances, follow-up disease control measures driven by high-fidelity testing and efficient contact tracking should be capable of effectively eradicating the few isolated disease resurgence centers. However, to reach these favorable configurations for the implementation of localized and targeted epidemic control, it is imperative to maintain the original social-distancing restrictions for at least a factor of three (better four) longer than it would have taken the unmitigated outbreak to reach its peak ($T \approx 3/a \ldots 6/a$ in our simulations) – for COVID-19 that would correspond to about two months. As is evident from our results for two-dimensional small-world networks that perhaps best represent human interactions, it is also absolutely crucial to severely limit all far-ranging links between groups to less than 5 % of the overall connections, during the disease outbreak.

Following this work, we have further looked at the effects of introducing the incubation period to the modeling of the epidemics spread (29). Our simulations of an extended Susceptible-Exposed-Infectious-Recovered (SEIR) compartmental model have shown that the incubation period sets a delay to the infection onset and induces a broadening of the infection curves in comparison to the SIR model.

## Data Availability

All data generated or analyzed during this study are included in this published article (and its Supplementary Information files).

# Acknowledgments

Research was sponsored by the U.S. Army Research Office and was accomplished under Grant No. W911NF-17-1-0156. The views and conclusions contained in this document are those of the authors and should not be interpreted as representing the official policies, either expressed or implied, of the Army Research Office or the U.S. Government. The U.S. Government is authorized to reproduce and distribute reprints for Government purposes notwithstanding any copyright notation herein. S.D. gratefully acknowledges a fellowship from the China Scholarship Council, Grant No. CSC 201806770029.

# Author Contributions

R.I.M. designed the algorithms for and ran Monte Carlo simulations on the regular square lattices.
S.D. constructed the small-network model variants and with S.S. ran numerical simulations on them.
P. generated and ran computer simulations on both random and scale-free network structures.
R.N. and H.Y. compared the stochastic simulation data generated on all architectures with numerical integrations of the deterministic SIR rate equations.
U.C.T. directed and supervised the entire research.
All authors contributed equally to the data analysis and interpretation, as well as on writing and editing the manuscript and supplementary material text and figures.

# Competing Interests Statement

The authors declare no competing interests.

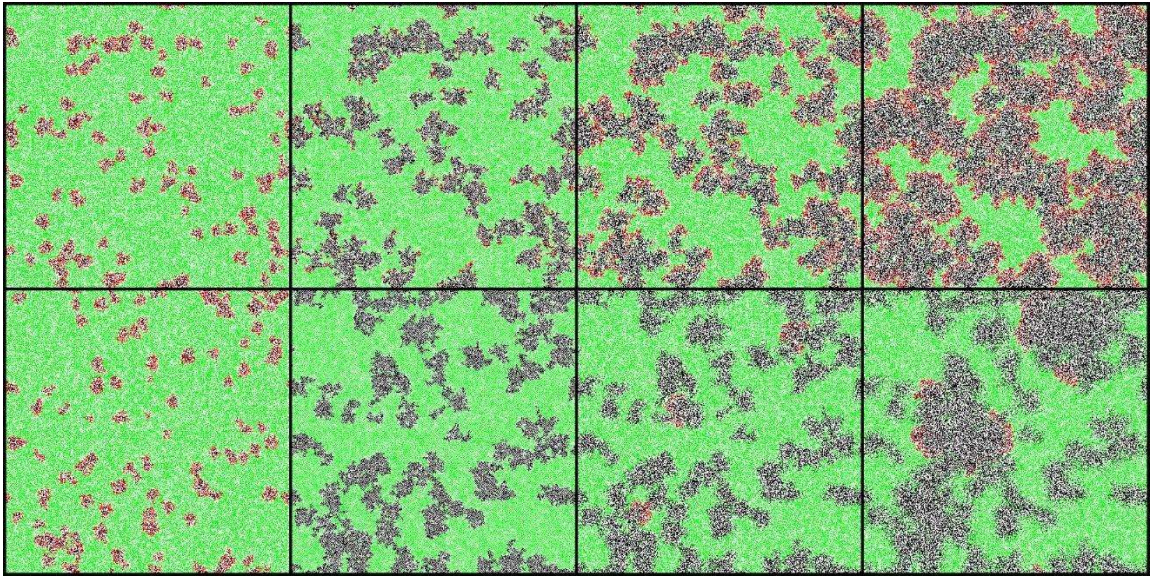

**Figure 1. Stochastic *SIR* model simulation snapshots on a square lattice (with periodic boundary conditions).**
'Social distancing' is turned on when the number of infective individuals reaches $I(t) = 0.1\,N$, and subsequently maintained for a duration $T = 22\,MCS = 2/a$ in the top, and $T = 110\,MCS = 10/a$ in the bottom row. The green color marker is used for susceptible individuals, while red indicates infected and black recovered (immune or deceased) individuals. The first snapshots (leftmost column) capture the instance when mitigation is implemented. The second column marks the time when social distancing is turned off after additional time $T$ has elapsed. The third and fourth columns show the ensuing spread of the disease. With extended social distancing duration $T$ (bottom row), the infection becomes more likely to be driven to extinction in confined contact regions. Hence the number of active outbreak centers decreases drastically, which could facilitate disease control through effective testing and tracking.

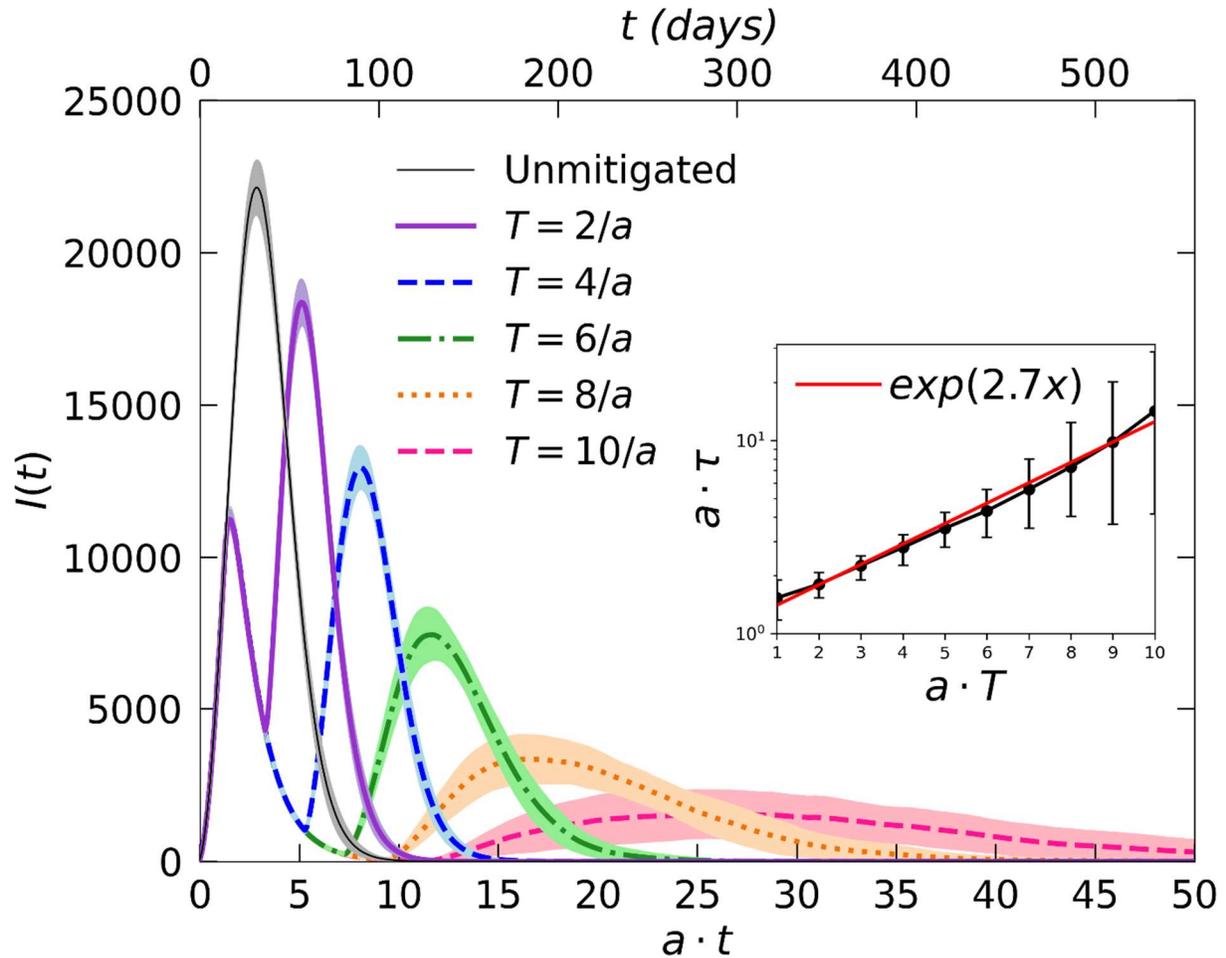

**Figure 2. Infection curves $I(t)$ for the stochastic *SIR* model on a square lattice.** The graphs compare the outbreak data obtained without any mitigation (grey) and with social distancing measures implemented for different durations $T$, as indicated. In all cases, social distancing is turned on once $I(t)$ reaches the set threshold of 10 % of the total population $N$. The resurgent outbreak is drastically reduced in both its intensity and growth rate as social distancing is maintained for longer time periods $T$. (The data for each curve were averaged over 100 independent realizations; the shading indicates statistical error estimates.)
Inset: time $\tau$ to reach the second peak following the end of the mitigation; the data indicate an exponential increase of $\tau$ with $T$.

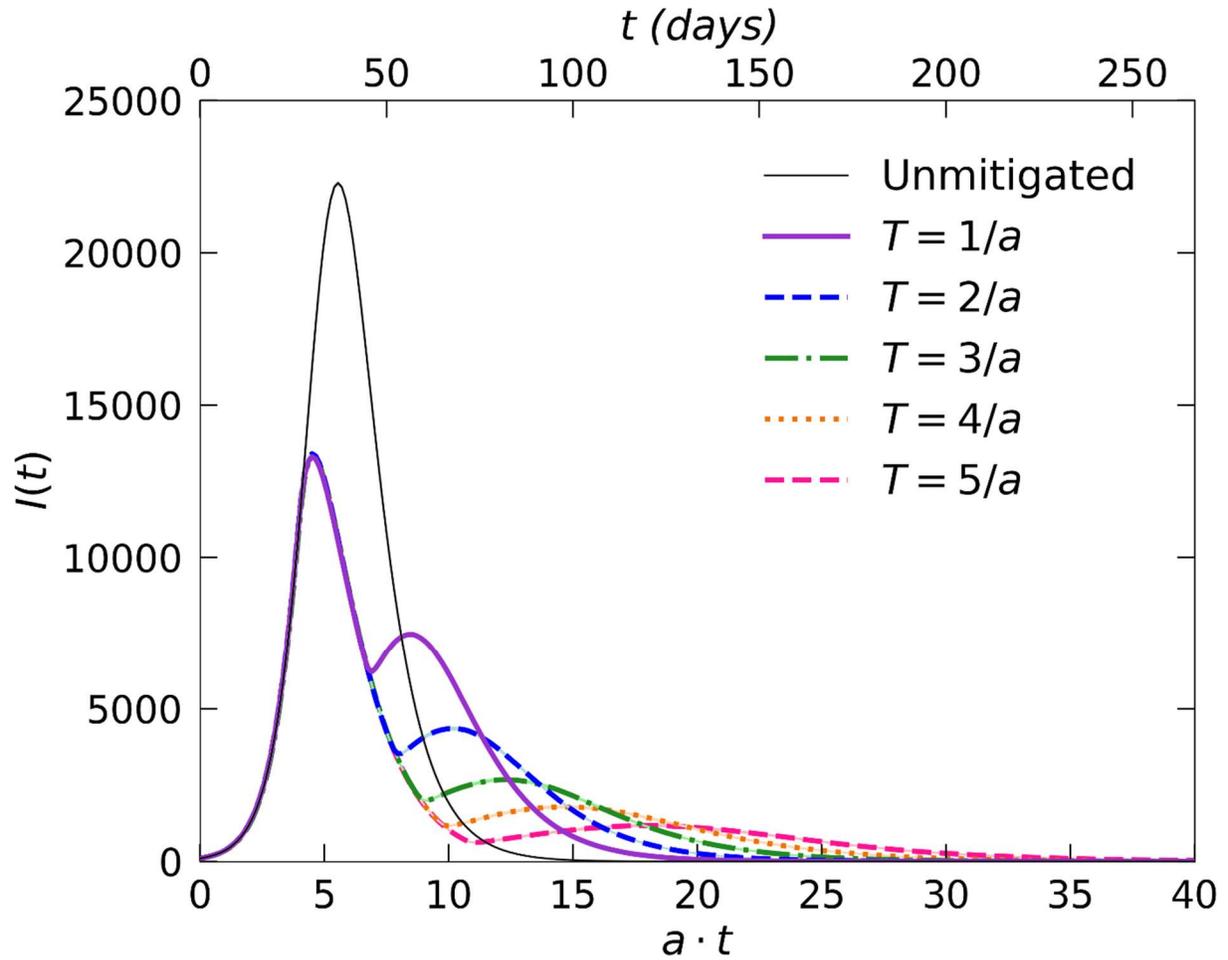

**Figure 3. Infection curves $I(t)$ from stochastic *SIR* model simulations on a two-dimensional Newman-Watts small-world network.** The graphs compare outbreak data without mitigation (grey) and for varying social-distancing intervention duration $T$ (as indicated), during which the probability of moving through long-distance connections was drastically reduced to $d_\varphi = 0.05$. (The data for each curve were averaged over 100 independent realizations.)

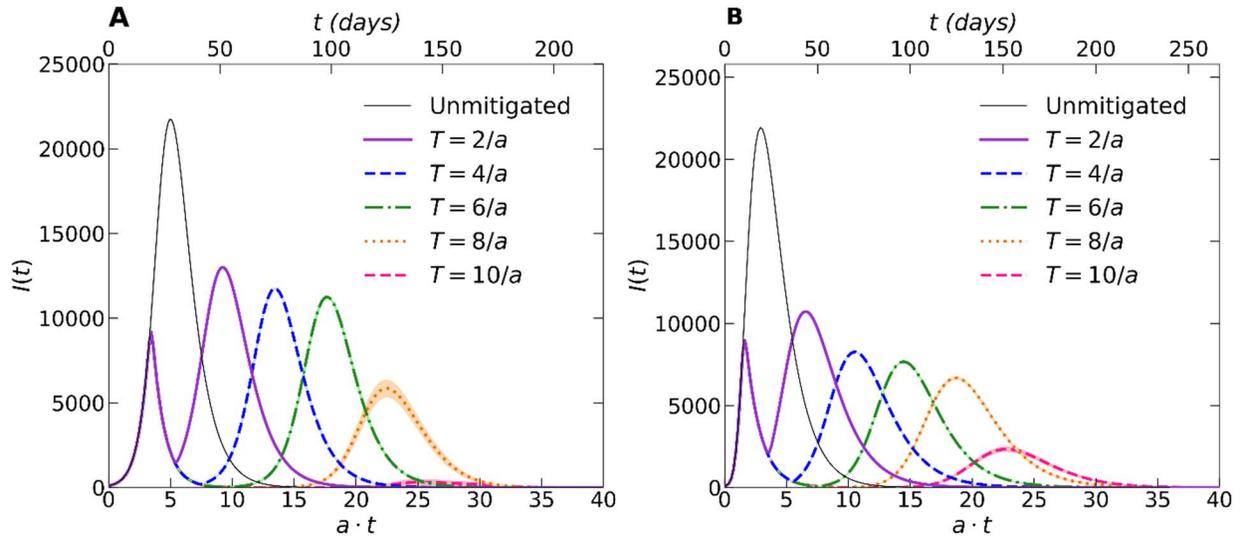

**Figure 4. Infection outbreak curves $I(t)$ from stochastic *SIR* model simulations.** (**A**) On a randomly connected network; (**B**) on a scale-free network with varying social-distancing intervention duration $T$. (The data for each curve were averaged over 100 independent realizations.)

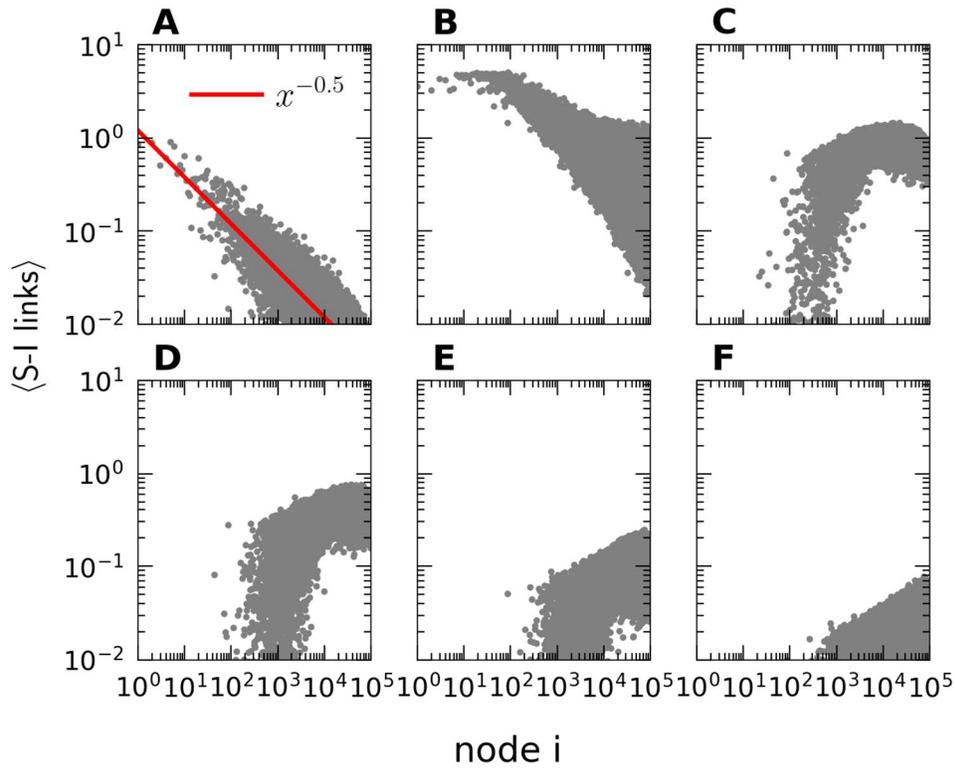

**Figure 5. Distribution of the mean number of susceptible-infectious (*SI*) connections for the nodes in scale-free contact networks**. (**A**) Network configuration before and (**B-F**) while the epidemic surge is spreading through the system. (**A**) at $t = 0$; (**B**) after 5 days; (**C**) after 10 days; (**D**) after 20 days, when the epidemic has reached its (unmitigated) peak; (**E**) after 30 days; (**F**) after 40 days, when only about 100 infectious individuals are left. (The data for each curve were averaged over 10,000 independent simulation runs.)

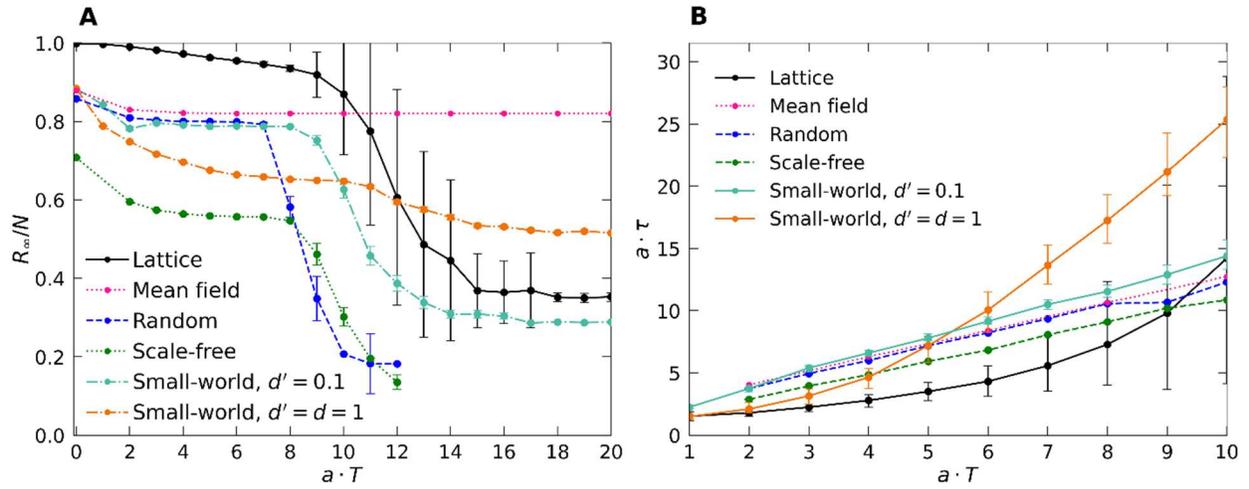

**Figure 6. Comparison of epidemic control measures through social distancing mitigation as functions of their duration $T$ on the various architectures**. (**A**) Recovered saturation fraction $R_\infty/N$; (**B**) time $\tau$ after release of the control measures until the infection resurgence peak is reached.

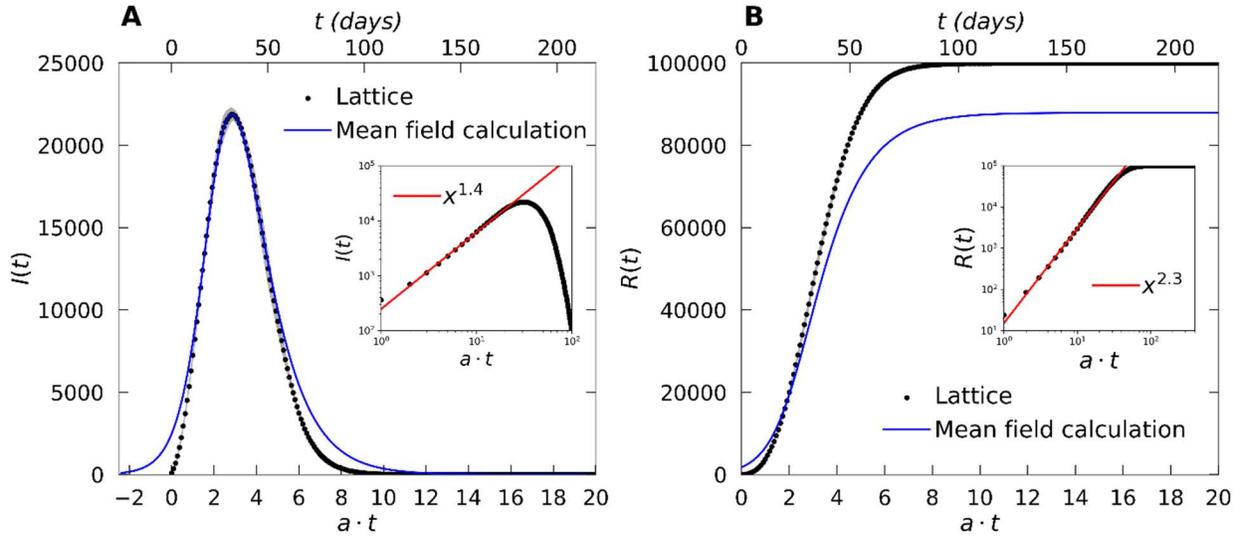

Figure S1. **Fit of infection curves from lattice simulations to the numerically integrated curves from the mean-field *SIR* rate equations.** (**A**) Infectious population $I(t)$; (**B**) recovered number of individuals $R(t)$. The insets illustrate the power law initial growth $I(t) \sim t^{1.4 \pm 0.1}$ and $R(t) \sim t^{2.3 \pm 0.1}$ for the lattice simulation data (averaged over 100 independent realizations).

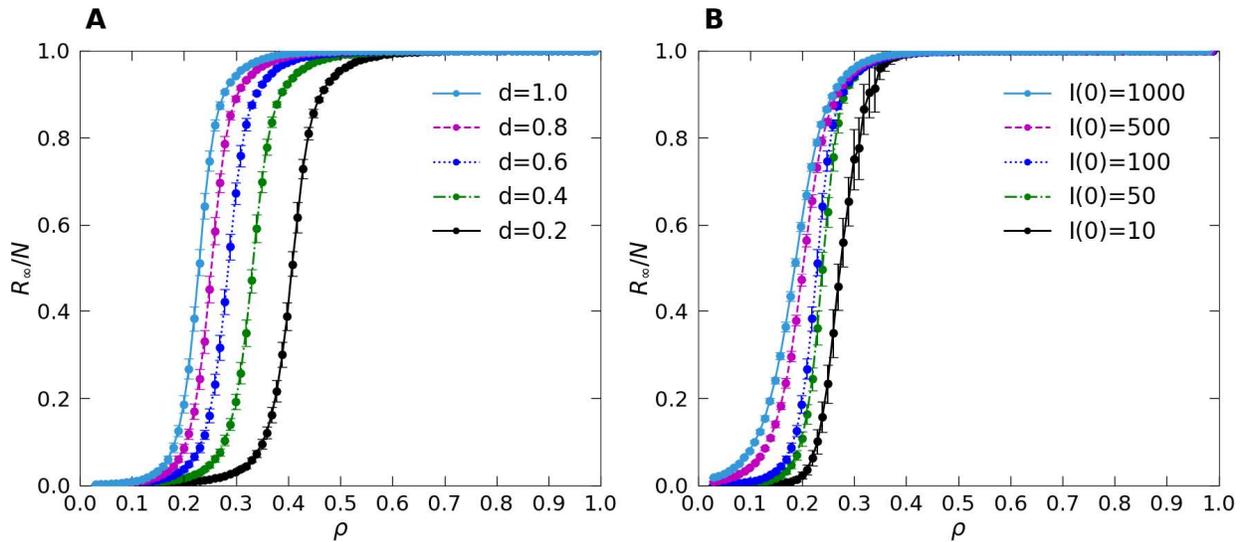

Figure S2: **Variation of the total fraction of recovered individuals $R_\infty/N$ with the lattice simulation parameters in stochastic *SIR* model simulations on a square lattice on the total density $\rho$.** (**A**) Data for different sets of nearest-neighbor hopping rates $d$; (**B**) for varying numbers of initially infected individuals $I(0)$. The graphs demonstrate the presence of a percolation-like epidemic threshold. (Each data point was averaged over 100 independent simulation runs.)

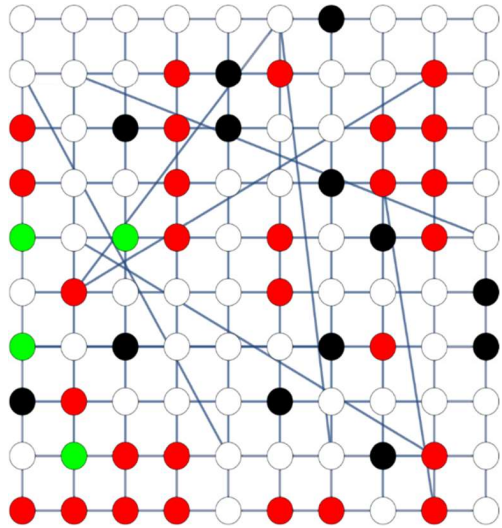

Figure S3: **Schematic construction of a two-dimensional Newman-Watts small-world network.** It is obtained from a regular square lattice through adding long-distance connections; shown is an *SIR* model configuration snapshot with empty, susceptible, infectious, and recovered states.

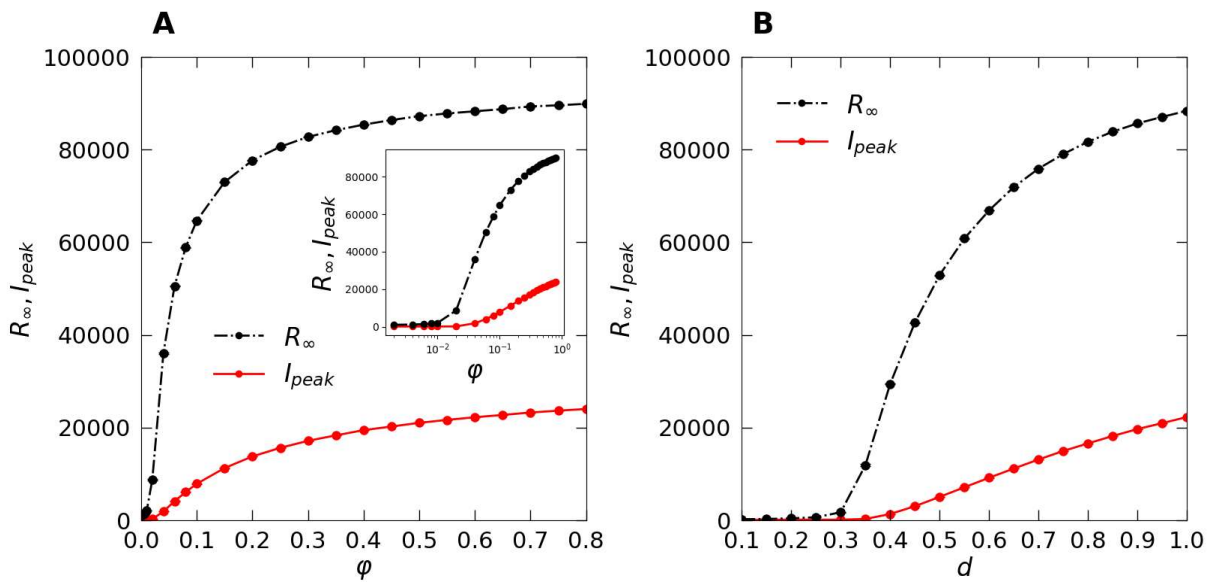

Figure S4: **Epidemic peak value and total recovered fraction $R_\infty$ and peak intensity $I_{peak}$.** (**A**) As functions of the fraction $\varphi$ of long-distance links with fixed $d = 1$ (inset: same data, with the $\varphi$ axis on a logarithmic scale); (**B**) as functions of the diffusivity $d$ with fixed $\varphi = 0.6$ for the SIR model implemented on a two-dimensional Newman-Watts small-world network. (Each data point was averaged over 100 independent realizations.)

# Supplemental Information

## Square lattices with diffusive spreading

On our regular square lattice with $L^2$ sites set on a two-dimensional torus, we implement the stochastic Susceptible-Infectious-Recovered (SIR) epidemic model with the following individual-based Monte Carlo algorithm:
1. Randomly distribute $N$ individuals on the lattice, subject to the restriction that each site may only contain at most one individual, and with period boundary conditions. Some small fraction of the individuals will initially be infectious, while the remainder of the population will be susceptible to the infection.
2. Perform random sequential updates $L^2$ times in one Monte Carlo step ($MCS$) by picking a lattice site at random, and then performing the following actions:
    a. If the selected site contains a susceptible $S$ or a recovered individual $R$, a hopping direction is picked randomly. If the adjacent lattice site in the hopping direction is empty, then the chosen individual is moved to that neighboring site with hopping probability $d$ that is related to a macroscopic diffusion rate.
    b. If the chosen lattice site contains an infectious individual $I$, it will first try to infect each susceptible nearest neighbor $S$ with a prescribed infection probability $r$. If this attempt is successful, the involved susceptible neighbor $S$ immediately changes its state to infected $I$. After the originally selected infected individual has repeated its infection attempts with all neighboring susceptibles $S$, it may reach the immune state $R$ with recovery probability $a$. Finally, this particular individual, whether still infectious or recovered, tries to hop in a randomly picked direction with probability $d$, provided the chosen adjacent lattice site is empty.
3. Repeat the procedures in item 2 for a preselected total number of Monte Carlo steps.

*Figure S1 about here.*

To determine the effective (coarse-grained) basic epidemic reproduction ratio $R_0$, we fit the infection curves to straightforward numerical integrations of the deterministic *SIR* rate equations $dS(t)/dt = -r\, S(t)\, I(t)/N$, $dI(t)/dt = r\, S(t)\, I(t)/N - a\, I(t)$, $dR(t)/dt = a\, I(t)$, and adjust the lattice simulation infection probability $r \approx 1.0$ and to a lesser extent, the recovery probability $a$ to finally match the targeted COVID-19 value $R_0 \approx 2.4$. We note that this slightly `renormalized' value for $a$ is subsequently utilized to set the time axis scale in the figures. On the mean-field level, initially $R_0 = (r/a)\, S(0)/N$, since all nodes are mutually connected. In spatial settings, $S(0)/N$ is to be replaced with the mean connectivity (i.e., the coordination number for a regular lattice) to susceptible individuals. The lattice simulation data is fitted with the mean-field result by matching two parameters: the maximum value and the half-peak width of the infectious population curve $I(t)$, see Figure S1. The lattice simulation curve digresses from the mean-field curves at low $I(t)$ values, far away from the peak region. In the lattice simulations, the initial rise of the infectious population curve exhibits power-law growths $I(t) \sim t^{1.4 \pm 0.1}$ and $R(t) \sim t^{2.3 \pm 0.1}$ in clear contrast with the simple exponential rise of the mean-field *SIR* curve as obtained from integrating the mean-field rate equations. We note that these are the standard critical exponents $\theta$ and $1 + \theta$ for the temporal growth of an active seed cluster near a continuous non-equilibrium phase transition to an absorbing extinction state *(11)*.

*Figures S2 about here.*

Figure S2 shows the dependence of the asymptotic number of recovered individuals $R_\infty$ on the density $\rho$ for various sets of hopping rates $d$ and initial infectious population values $I(0)$. These data indicate the existence of a well-defined epidemic threshold, i.e., a percolation-like sharp transition from a state when only a tiny fraction of individuals is infected, to the epidemic state wherein the infection spreads over the entire population *(11)*. As one would expect, this critical point depends only on the ratio $a/d$ of the recovery and hopping rates. Varying the lattice simulation parameters just shifts the location of the epidemic threshold. Once the model parameters are set in the epidemic spreading regime, the system's qualitative behavior is thus generic and robust, and only weakly depends on precise parameter settings.

**Two-dimensional small-world networks**

*Figure S3 about here.*

For our two-dimensional small-world network, whose construction is schematically depicted in Figure S3, we employ a similar Monte Carlo algorithm as described above; the essential difference is that individuals may now move to adjacent nearest-neighbor as well as to distant lattice sites along the pre-set `short-cut' links. Figure S4A demonstrates (for fixed diffusivity $d = 1$) that as function of the fraction $\varphi$ of long-distance links in a two-dimensional small-world network, the epidemic threshold resides quite close to zero: The presence of a mere few `short-cuts' in the lattice already implies a substantial population mixing. The inset, where the $\varphi$ axis is scaled logarithmically, indicates that sizeable outbreaks begin for $\varphi \geq 0.05$. Figure S4B similarly shows the outbreak dependence on the diffusion rate $d$ (here for $\varphi = 0.6$), with the threshold for epidemic spreading observed at $d \approx 0.3$. Evidently, prevention of disease outbreaks in this architecture requires that both mobility and the presence of far-ranging connections be stringently curtailed.

*Figure S4 about here.*

**Random and scale-free contact networks**

For both the randomly connected and scale-free contact networks, we employ the Gillespie or dynamical Monte Carlo algorithm, which allows for efficient numerical simulations of Markovian stochastic processes. It consists of these subsequent steps:
1. Initially, few nodes are assigned to be infected $I$, while all other nodes are set in the susceptible state $S$. Each susceptible node $S$ is characterized by a certain number of active links that are connected to infected nodes $I$.
2. We then determine the rate at which each infected node $I$ will recover, and at which each susceptible node $S$ with a non-zero number of active links becomes infected. From these we infer the total event rate $r_{tot}$.
3. Based on this total rate $r_{tot}$, we select the waiting time until the next event occurs from an exponential distribution with mean $r_{tot}$.
4. We then select any permissible event with a probability proportional to its rate, update the status of each node, and repeat these processes for the desired total number of iterations.